\documentclass[epj,pacs]{svjour}

\usepackage{latexsym}
\usepackage{indentfirst}
\usepackage{amsmath}
\usepackage{amssymb}
\usepackage{color}
\usepackage{graphicx}
\usepackage{bm}
\usepackage{maybemath}
\usepackage{revsymb4-1}

\newcommand{\dd}{\mathrm{d}}
\newcommand{\ee}{\mathrm{e}}
\newcommand{\ii}{\mathrm{i}}
\newcommand{\calA}{{\cal A}}

\journalname{EPJ C}

\begin{document}

\sloppy

\title{Compton Upconversion of Twisted Photons:\\
Backscattering of Particles with Non-Planar Wave Functions}

\titlerunning{Compton Upconversion of Twisted Photons}

\author{U. D. Jentschura \inst{1,2} \and
V. G. Serbo \inst{2,3,}\thanks{e-mail: serbo@math.nsc.ru}}

\authorrunning{Jentschura and Serbo}

\institute{Department of Physics, Missouri University of Science
and Technology, Rolla, Missouri 65409-0640, USA \and
Institut f\"ur Theoretische Physik,
Universit\"{a}t Heidelberg,
Philosophenweg 16, 69120 Heidelberg, Germany \and
Novosibirsk State University,
Pirogova 2, 630090, Novosibirsk, Russia}

\date{Received: 26--OCT--2010}

\abstract{Twisted photons are not plane waves, but superpositions of plane
waves with a defined projection $\hbar m$ of the orbital angular momentum onto
the propagation axis ($m$ is integer and may attain values $m \gg 1$). Here, we
describe in detail the possibility to produce high-energy twisted photons by
backward Compton scattering of twisted laser photons on ultra-relativistic
electrons with a Lorentz-factor $\gamma=E/(m_ec^2) \gg 1$. When a twisted laser
photon with the energy $\hbar \omega \sim 1\,{\rm eV}$ performs a collision
with an electron and scatters backward, the final twisted photon conserves the
angular momentum $m$, but its energy $\hbar\omega'$ is increased considerably:
$\omega'/\omega=4\gamma^2/(1+x)$, where $x=4E \, \hbar\omega/(m_e \,c^2)^2$. The
$S$ matrix formalism for the description of scattering processes is
particularly simple for plane waves with definite 4-momenta. However, in the
considered case, this formalism must be enhanced because the quantum state of
twisted particles cannot be reduced to plane waves.  This implies that the
usual notion of a cross section is inapplicable, and we introduce and calculate
an averaged cross section for a quantitative description of the process. The
energetic upconversion of twisted photons may be of interest for experiments
with the excitation and disintegration of atoms and nuclei, and for studying
the photo-effect and pair production off nuclei in previously unexplored
regimes.}

\maketitle


%
%
\section{Introduction}
\label{intro}

Scattering processes lie at the heart of modern physics and have been studied
in detail at the tree- and loop-level for particles with well-defined four
momenta. In particular, the well-known Feynman rules of quantum
electrodynamics~\cite{BeLiPi1982vol4,ItZu1980} apply to the scattering of
planar waves and cannot be readily applied to the scattering of particles with
more complex wave functions.  Consequently, the literature is more scarce when
it comes to the scattering of quantal particles described by non-planar waves.

The related questions are far from being academic. E.g.,
a description using non-planar waves is necessary in the case
of single photon bremsstrahlung discovered in
experiments on the $e^+ e^-$ collider 
VEPP-4~\cite{BlEtAl1982,BaKaSt1982,BuDeMisc} and
then on the $ep$ collider HERA~\cite{Pi1995}. In these experiments, a
remarkable deviation of the measured bremsstrahlung photons from standard
calculational methods has been observed. The decrease in the number of 
observed photons can be 
explained by the fact that large impact parameters 
(of the two bunches relative to each other) give the essential
contributions to the cross section. These parameters are larger by several
orders of magnitude than the transverse beam size. In that case, the standard
definitions for the cross section and the number of events become invalid. In
particular, it is possible to calculate the $S$ matrix element as a
superposition of ``planar'' scattering processes, but the normalization of the
flux of incoming particles still constitutes a problem in the calculation of
the modified cross section (beyond the $S$ matrix element). 
Modified calculational schemes for the description of 
particle production in the interaction of two bunches 
have to be employed (for details, see the review~\cite{KoSeSc1992}).
In this scheme, the colliding bunches are represented as wave packets, and
quantum distribution functions are used. The modified definitions of the cross
section and the number of events contain the features of ``non--locality'' and
``interference.''

An analogous problem is studied here for Compton backscattering of so-called
``twisted photons.'' These are defined superpositions of plane waves and have
some interesting physical properties~\cite{FAAlPa1992,PaCoAl2004}, such as
wavefronts that rotate about the propagation axis and Poynting vectors that
look like corkscrews (see Fig.~1 of Ref.~\cite{MaVaWeZe2001}).  Also, twisted
photons have a defined projection of the orbital angular momentum  $\hbar m$
onto the propagation axis~\cite{Gr2003,FAAlPa2008} which may be quite large,
$m\sim 100$. Experiments demonstrate that micron-sized Teflon, calcite and
other micron-sized ``particles'' start to rotate after absorbing such
photons~\cite{HeFrHeRD1995,SiDhPa1997,FrNiHeRD1998,ONEtAl2002,SiDhAlPa2003}.
The observation of orbital angular momentum of light scattered by black holes
could be very instructive, as pointed out in Ref.~\cite{Ha2003}. Twisted laser
photons may be created from usual laser beams by means of numerically computed
holograms.  Alternative generation mechanisms (in the visible and infrared part
of the optical spectrum) have recently been discussed in
Refs.~\cite{ArBa2000,BaTa2003}.

The electromagnetic vector potential describing a twisted photon
state adds the orbital angular momentum of the photon to the 
spin angular momentum of the vector (spin-$1$) field.
In some sense, the twisted wave function interpolates between the 
plane-wave vector potential of the form
$e_{\vec k \, \Lambda} \, \exp(\ii \vec k \vec r)$ 
and the photon vector potential
described by a vector spherical harmonic ${\vec Y}_{JLM}(\theta, \phi)$.
Indeed, a plane-wave photon whose vector 
potential is proportional to 
$e_{\vec k \Lambda} \, \exp(\ii k_z \, z)$,
describes a photon propagating in the $z$ direction.
It has zero expectation value for the 
projection of the angular momentum $\check L_z$ onto the 
propagation axis ($z$ axis). A photon described by 
a vector spherical harmonic ${\vec Y}_{JLM}(\theta, \phi)$
fulfills $\check{\vec J}^2 \, {\vec Y}_{JLM}(\theta, \phi) =
J\,(J+1) \, {\vec Y}_{JLM}(\theta, \phi)$ and
$\check{J}_z {\vec Y}_{JLM}(\theta, \phi) =
M \, {\vec Y}_{JLM}(\theta, \phi)$,
where $\check{\vec J} = \check{\vec L} + \check{\vec S}$ is the total angular momentum
(orbital plus spin) of the photon.
However, a photon described by a vector spherical harmonic does not
have a defined propagation direction.

Twisted photons are rather interesting objects, as they combine, in some sense,
the properties of plane-wave photons and those described by vector spherical
harmonics: Namely, they have a defined propagation direction 
(which we choose to be the $z$ axis, here) and still, large
angular momentum projections onto that same propagation axis.  In constructing
vector spherical harmonics, one adds the orbital angular momentum from
the spherical harmonics to the spin angular momentum, using Clebsch--Gordan
coeffcients~\cite{VaMoKh1988}.  However, one can also add the orbital angular
momentum to the spin angular momentum via a conical momentum spread (in
momentum space) multiplied by an angle-dependent phase, or by a Bessel function
in the radial variable (in coordinate space). This leads to the twisted states,
which are the subject of the current paper.

All experiments performed with twisted photons so far have been in the range of
visible light, i.e., with a photon energy of the order of $1\,{\rm eV}$.  In
our recent paper~\cite{JeSe2011}, we have shown that it is possible to
upconvert the frequency of a twisted photon using Compton backscattering, from
an energy of the order of $1\, {\rm eV}$ to an energy in the GeV~range. Here,
we present the derivation in more detail, and we also address the question of
how to convert the result for the $S$ matrix element to a generalized cross
section. This is nontrivial for the current case because the initial and final
photons are described as twisted states (rather than plane waves).

This paper is organized as follows. In Sec.~\ref{scalartwisted}, we present
basic formulas pertaining to a twisted state of a scalar particle, whereas the
full vector particle content of a twisted photon is investigated in
Sec.~\ref{photontwisted}. The Compton scattering is recalled in
Sec.~\ref{comptonplane} for a plane-wave photon, whereas the same effect is
studied for backscattered twisted photons in Sec.~\ref{comptontwisted}.
Details of the conversion of the $S$ matrix element to a generalized cross
section are discussed in Sec.~\ref{avcrossX} and Appendices~\ref{appa}
and~\ref{appb}.  Finally, conclusions are reserved for Sec.~\ref{conclu}.
Relativistic Gaussian units with $c=1$, $\hbar = 1$, $\alpha\approx 1/137$, and
$\epsilon_0 = (4 \pi)^{-1}$ are used throughout the article. We write the
electron mass as $m_e$ and denote the scalar product of 4-vectors
$k=(\omega,\,{\bm k})$ and $p=(E,\,{\bm p})$ by a dot, i.e.,~$k\cdot p= \omega
\, E - {\bm k} {\bm p}$, where ${\bm k} {\bm p}$ is the scalar product of
3-vectors.

%
%
\section{Quantum description of twisted states}

%
%
\subsection{Twisted scalar particle}
\label{scalartwisted}

\begin{figure}
\includegraphics[width=0.99\linewidth]{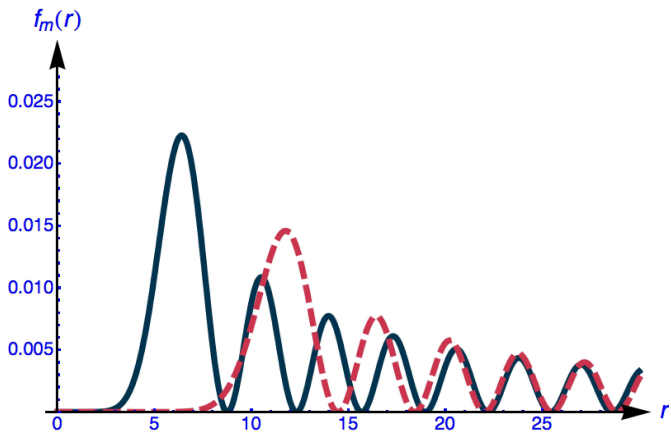}
\caption{\label{fig1} (Color online.) Plot of the radial
probability density $f_m(r) = |\psi_{\varkappa m}(r,\varphi)|^2$
for $m = 5$ (solid line) and $m=10$ (dashed line) at
$\varkappa=1$. The central minimum of the wave function and the
maximum at $r\sim m$ are clearly visible.}
\end{figure}

We first recall that the usual plane-wave state of a scalar
particle with mass equal to  zero has a defined 3-momentum ${\bm k}$,
energy $\omega=|{\bm k}|$ and is described by a wave function of
the form
\begin{equation}
\Psi_{\bm k}(t,{\bm r}) = \;
\frac{\ee^{-\ii (\omega \, t-{\bm k}  {\bm r})}}{\sqrt{2\omega}} \,,
\end{equation}
with the normalization condition
\begin{equation}
\label{norm}
\int \Psi^*_{{\bm k}'}(t,{\bm r})\,\Psi_{\bm k}(t,{\bm r})
\,\dd^3r=\frac{(2\pi)^3 \delta({\bm k}-{\bm k}')}{2\omega} \,.
\end{equation}
A twisted scalar particle with vanishing mass has the following
quantum numbers: longitudinal momentum $k_z$, absolute value of
the transverse momentum $\varkappa$, energy
\begin{equation}
\omega=|{\bm k}|= \sqrt{\varkappa^2+k_z^2}\,,
\end{equation}
and projection $m$ of the orbital angular momentum onto the $z$
axis. In cylindrical coordinates $r$, $\varphi_r$, and $z$, this
state is described by the wave function $\Psi_{\varkappa m k_z
}(t,{\bm r})\equiv\Psi_{\varkappa m k_z}(r,\varphi_r,z,t)$ which
satisfies the Klein-Fock-Gordon equation (with mass equal to
zero),
\begin{equation}
\partial_\mu\partial^\mu\,\Psi_{\varkappa m k_z }(t,{\bm r})=0 \,,
\end{equation}
and it is an eigenfunction of the $z$ projection of the momentum
$\check{p}_z=-\ii \partial/\partial z$ and of the orbital angular
momentum $\check{L}_z=-\ii\partial/\partial\varphi_r$,
\begin{equation}
\check{p}_z\,\Psi_{\varkappa m k_z }= k_z\,\,\Psi_{\varkappa m
k_z}\,,\;\check{L}_z\,\Psi_{\varkappa m k_z}=m\,\Psi_{\varkappa m
k_z }\,.
\end{equation}
Its evident form is
\begin{align}
\label{defPsi} \Psi_{\varkappa m k_z}(r,\varphi_r,z,t) =& \;
\frac{\ee^{-\ii (\omega t - k_z z)}}{\sqrt{2\omega}}\;
\psi_{\varkappa m}(r,\varphi_r) \,,
\nonumber\\[2ex]
\psi_{\varkappa m}(r,\varphi_r) =& \; \frac{\ee^{\ii
m\varphi_r}}{\sqrt{2\pi}}\, \sqrt{\varkappa}\, J_m(\varkappa \;
r)\,,
\end{align}
where $J_m(x)$ is the Bessel function
\begin{equation}
J_m(x)=\frac{1}{2\pi}\, \int_0^{2\pi} \ee^{\ii
(m\varphi-x\sin{\varphi)}}\, \dd\varphi \,.
\end{equation}
In Fig.~\ref{fig1}, we present the dependence of the square of the 
absolute
value of $\psi_{\varkappa m}(r,\varphi_r)$ on $r$ for different
values of $m$. For small $r\ll 1/\varkappa$, this function is
of order $r^m$,
\begin{equation}
|\psi_{\varkappa m}(r,\varphi_r)| \approx \sqrt{\frac{\varkappa}{2
\pi}} \, \frac{(\varkappa r)^m }{2^m \, m!} \,,
\end{equation}
has a maximum at $r\sim m/\varkappa$ and then drops according to the
familiar asymptotics of the Bessel function,
\begin{equation}
\psi_{\varkappa m}(r,\varphi_r) \approx \frac{\ee^{\ii
m\varphi_r}}{\pi \sqrt{r}}\, \cos{\left(\varkappa \, r - \frac{m
\, \pi}{2}- \frac{\pi}{4} \right)}\,,
\end{equation}
at large values $r\gg 1/\varkappa$.

The function $\psi_{\varkappa m}(r,\varphi)$ may be expressed as a
superposition of plane waves in the $xy$ plane
($\bm k_\perp \, {\bm e}_z = 0$)
\begin{equation}
\label{Psi} \psi_{\varkappa m}(r,\varphi) = \int a_{\varkappa
m}({\bm k}_\perp)\, \ee^{\ii{\bm k}_\perp  {\bm r}}\,
\frac{\dd^2k_\perp}{(2\pi)^2}\,,
\end{equation}
where the Fourier amplitude $a_{\varkappa m}({\bm k}_\perp)$ is
concentrated on the circle with $k_\perp \equiv |{\bm k}_\perp| =
\varkappa$,
\begin{equation}
a_{\varkappa m}({\bm k}_\perp) = (-\ii)^m \; \ee^{\ii m\varphi_k}
\; \sqrt{\frac{2\pi}{\varkappa}}\, \delta(k_\perp-\varkappa) \,.
 \label{Ft}
\end{equation}
Therefore, the function $\Psi_{\varkappa m k_z }(r,\varphi_r,z,t)$
can be regarded as a superposition of plane waves with defined
longitudinal momentum $k_z$, absolute value of transverse
momentum  $\varkappa$, energy $\omega= \sqrt{\varkappa^2+k_z^2}$
and different directions of the vector ${\bm k}_\perp$ given by
the angle $\varphi_k$.

\begin{figure}
\begin{center}
\includegraphics[width=1.0\linewidth]{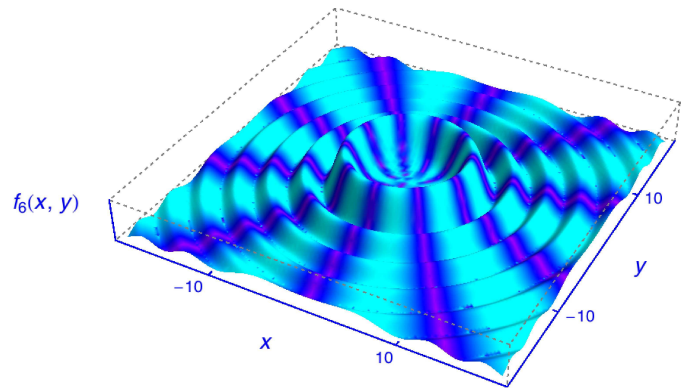}\\
(a)\\
\vspace*{0.5cm}
\includegraphics[width=1.0\linewidth]{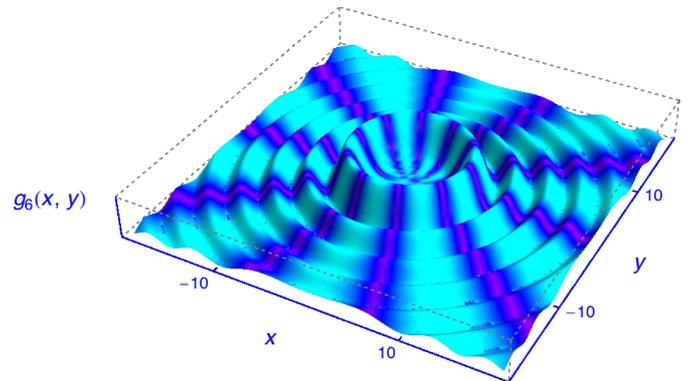}\\
(b)
\end{center}
\caption{\label{fig2} (Color online.) Illustration of the quantum
vector potential  ${\calA}^{\mu}_{\varkappa m k_z \Lambda}$ of a
twisted photon. The plot displays 
$f_m(x,y) = |{\calA}^{\mu=1}_{\varkappa m k_z \Lambda}(0, x,y,0)|^2$ 
and $g_m(x,y) = |{\calA}^{\mu=2}_{\varkappa m k_z \Lambda}(0, x,y,0)|^2$ 
as a function of $x$ and $y$. 
The parameters are $\varkappa = 1$, $m = 6$,
$k_z = \sqrt{48}$, 
$\omega = \sqrt{\varkappa^2 + k_z^2} = 7$, and
$\Lambda = 1$. The complex phase of the vector potential 
(modulo~$\pi$) is
indicated by the variation of the color/grayscale of the wave
function. Panels~(a) and ~(b) illustrate that the 
$x$ and $y$ components ($\mu=1$ versus $\mu= 2$) of the twisted photon
vector potentials are phase shifted with respect
to each other.}
\end{figure}

%
%
\subsection{Twisted photon}
\label{photontwisted}

The wave function of a twisted photon (vector particle) can be
constructed as a generalization of the scalar wave function. We
start from the plane-wave photon state with a defined
4-momentum $k=(\omega, {\bm k})$ and helicity $\Lambda=\pm 1$,
\begin{subequations}
\begin{align}
\label{photon} 
A^\mu_{k\Lambda}(t, \bm r) =& \;
\sqrt{4\pi}\, e^\mu_{k\Lambda}\,
\frac{\ee^{-\ii (\omega \, t-{\bm k}  {\bm r})}}%
{\sqrt{2\omega}}\,,
\\[2ex]
e_{k\Lambda} \cdot k =& \; 0\,, \quad e^*_{k\Lambda} \cdot
e_{k\Lambda'} = -\delta_{\Lambda \Lambda'} \,,
\end{align}
\end{subequations}
where $e^\mu_{k\Lambda}$ is the polarization four-vector of the
photon. The twisted photon vector potential
\begin{align}
\label{twistedwave} &
{\calA}^\mu_{\varkappa m k_z
\Lambda}(r,\varphi_r,z,t) = \int a_{\varkappa m}({\bm k}_\perp) \,
A^\mu_{k\Lambda}(t, \bm r) \, \frac{\dd^2 k_\perp}{(2\pi)^2}
\\[2ex]
& = (-\ii)^m \sqrt{2\pi\varkappa} \int\limits_0^{2\pi}\dd\varphi_k
\int\limits_0^{\infty} \dd k_\perp\, \delta(k_\perp-\varkappa)\,
\frac{e^{\ii m\varphi_k}}{(2\pi)^2}\,A^\mu_{k\Lambda} (t, \bm r)
\nonumber
\end{align}
is given as a two-fold integral over the perpendicular components
${\bm k}_\perp = (k_x, k_y, 0)$ of the wave vector $\bm k =
(k_x,\,k_y,\, k_z)$. Using the well-known identity
\begin{equation}
\int_0^\infty \,  J_m(\varkappa \, x) \, J_m(\varkappa' \,
x)\,x\,\dd x = \frac{1}{\varkappa} \, \delta(\varkappa -
\varkappa')\,,
 \label{identJJ}
\end{equation}
it is not difficult to prove that this function satisfies the
normalization condition [compare with Eq.~\eqref{norm}]
\begin{align}
& \int \left( \calA^*_{\varkappa' m' k'_z \Lambda'} \right)_\mu(t,{\vec r})\,
\calA^\mu_{\varkappa m k_z \Lambda}(t,{\bm r}) \,\dd^3 r
\nonumber\\
&=-4\pi \delta_{\Lambda \Lambda'}\,\frac{2\pi
\delta(k_z-k'_z)}{2\omega}
\delta_{mm'}\delta(\varkappa-\varkappa').
\end{align}
We would like to stress that the orthogonal functions
$\calA^\mu_{\varkappa m k_z \Lambda}(t,{\bm r})$  for different values
of $\varkappa, m, k_z, \Lambda$ but fixed
$z$ axis constitute {\em a complete set of functions} and can be used
for the description of initial as well as final twisted photons.

The polarization vector $e_{k\Lambda}$ depends on the azimuth
$\varphi_k$ as $\exp( \ii \, \ell \varphi_k)$ with $\ell = 0,\,\pm 1$
depending on the helicity [see Eq.~(\ref{e1decomp}) below]. In
view of the identity
\begin{equation}
\label{int} \int \ee^{\ii \ell \varphi_k}\, a_{\varkappa m}({\bm
k}_\perp)\, \ee^{\ii {\bm k}_\perp  {\bm r}}\, \frac{\dd^2
k_\perp}{(2\pi)^2} = \ii^\ell\,\psi_{\varkappa, m+\ell}(r,\varphi_r)\,,
\end{equation}
the vector field $\calA^\mu_{\varkappa m k_z
\Lambda}(r,\varphi_r,z,t)$ describes a photon state with defined
$k_z$, absolute value of the transverse momentum $|{\bm
k}_\perp|=\varkappa$, energy $\omega= \sqrt{\varkappa^2+k_z^2}$
and projection of the orbital angular momentum on the $z$
axis equal to $m-1,\,m,\,m+1$ [see also Eqs.~\eqref{psi}---\eqref{Amu}
below]. Strictly speaking, this state is not a photon state with a
defined value of $\check{L}_z$. However, for large $m$, the
restriction to $(m-1,m,m+1)$ means that the twisted state is a
state with a very restricted angular momentum projection
distribution about the central value equal to $m$. The
representation~\eqref{int} is very convenient as it allows us to
considerably simplify the analytic calculations. We call such a
state a twisted $m$ photon (see Fig.~\ref{fig2}) and denote it as
$|\varkappa, m, k_z, \Lambda\rangle$.

\begin{figure}
\begin{center}
\includegraphics[width=0.99\linewidth,angle=0]{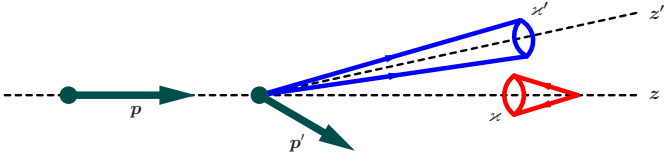}
 \caption{\label{fig3}
(Color online.) Schematic view of the scattering geometry. The
incoming twisted photon with transverse momentum spread
$\varkappa$ is scattered by a counterpropagating electron with
momentum $p$, into a state propagating along an arbitrary $z'$
axis with transverse momentum spread $\varkappa'$, with the
electron assuming momentum $p'$. For the specific case of
strict backward Compton scattering, the $z$  and $z'$ axes and the
vectors ${\bm p}$ and ${\bm p}'$ are all collinear.}
\end{center}
\end{figure}

The usual $S$ matrix element for plane-wave (PW) Compton
scattering involves an electron being scattered from the state
$|p, \lambda \rangle$ with 4-momentum $p$ and helicity
$\lambda=\pm \tfrac12$ to a state $|p', \lambda' \rangle$ and a
photon being scattered from the state $|k, \Lambda\rangle$ to the
state $|k', \Lambda' \rangle$,
\begin{equation}
\label{Sfi} S^{\rm (PW)}_{fi} \equiv
\langle k',\Lambda', p',\lambda'| S |k, \Lambda, p,\lambda\rangle \,.
\end{equation}
For head-on collisions, the vectors
${\bm p}=(0,0, p_z)$ and ${\bm k}=(0,0,-\omega)$
are anti-parallel.

Let us consider the Compton effect for the case when an initial
plane-wave electron in the state $|p, \lambda \rangle$ performs a
head-on collision with an initial twisted $m$ photon $|\varkappa,
m, k_z, \Lambda\rangle$ propagating along the $(-z)$ axis. In the
final state, there is a plane-wave electron $|p', \lambda'
\rangle$  and a final twisted $m'$ photon $|\varkappa', m', k'_z,
\Lambda' \rangle$ propagating along the $z'$ axis (a schematic
view of the initial and final states is given in Fig.~\ref{fig3}).
As noted above, we can choose the $z'$ axis along an arbitrary
direction, but below we restrict the discussion to the particular
scattering geometry where the axes $z'$ and $z$ are naturally
defined as being collinear, namely, the strict backscattering
geometry. (Moreover, even in a general case it is convenient to
choose the $z'$ axis along the $z$ axis, leading to a
potential simplification of the calculation.)
In view of Eq.~\eqref{twistedwave},
the $S$ matrix element for such a scattering,
\begin{equation}
S^{\rm (TW)}_{fi} \equiv \langle \varkappa', m',k'_z,\lambda'; 
p', \lambda' | S
| \varkappa, m, k_z, \Lambda; p, \lambda \rangle,
\end{equation}
needs to be integrated as follows,
\begin{align}
\label{convol} S^{\rm (TW)}_{fi} \equiv & \;
\int \frac{\dd^2k_\perp}{(2\pi)^2}\,\frac{\dd^2k'_\perp}{(2\pi)^2} \,
a^*_{\varkappa' m'}({\bm k}'_\perp) \;
\\[2ex]
& \times
\langle k',\Lambda', p',\lambda'| S |k, \Lambda, p,\lambda\rangle \,
a_{\varkappa m}({\bm k}_\perp) 
\nonumber\\[2ex]
=& \; \int
\frac{\dd^2k_\perp}{(2\pi)^2}\,\frac{\dd^2k'_\perp}{(2\pi)^2} \,
a^*_{\varkappa' m'}({\bm k}'_\perp) \; S^{\rm (PWC)}_{fi} \;
a_{\varkappa m}({\bm k}_\perp) \,,
\nonumber
\end{align}
where by PWC we denote the scattering matrix elements for the
plane-wave component of the twisted photons, with 4-vector
components $k = (\omega, \bm k_\perp, k_z)$ and  $k' = (\omega',
\bm k'_\perp, k_z')$.

Based on Eq.~\eqref{Ft}, we conclude that the integration in
Eq.~\eqref{convol} is determined by the dependence of the matrix
element $S^{\rm (PWC)}_{fi}$ on the azimuthal angles $\varphi_k$
and $\varphi'_{k}$ of the vectors ${\bm k}$ and ${\bm k}'$. A
numerical integration of Eq.~\eqref{convol} then leads to
predictions for arbitrary scattering angle of the final electron.
In this paper we consider in detail the important case of {\it
strict backward Compton scattering} when the scattering angle of
the final electron equals zero and the vector ${\bm p}'$ is
directed along the $z$ axis. Such a choice is determined mainly by
two reasons. First of all, for usual Compton scattering on
ultra-relativistic unpolarized electrons, precisely the backward
scattering has the largest probability (see Fig.~\ref{fig4}
below). Second, the matrix element
$S^{\rm (PWC)}_{fi}$ does not depend on the azimuthal angles
$\varphi_k$ and $\varphi'_{k}$, and therefore, this case allows
for a simple and transparent treatment with analytical
calculations for usual as well as for twisted photons.

%
\section{Compton scattering of plane-wave photons}
\label{comptonplane}

%
%
\subsection{General formulas}

In principle, Compton backscattering is an established method for
the creation of high-energy photons and used successfully in
various application areas from the study of photo-nuclear
reactions~\cite{NeTuSh2004} to colliding photon
beams of high energy~\cite{BaEtAl2004,Se2006appb}.
Let us consider the collision of an ultra-relativistic electron
with four momentum
\begin{equation}
p=(E,0,0,vE), \;\; v=\frac{|{\bm p}|} {E}, \;\;
\gamma=\frac{E}{m_e} \,,
\end{equation}
whose spatial momentum component points strictly upward,
and a photon of energy $\omega$ and three-momentum
\begin{equation}
{\bm k} = \omega \; (\sin{\alpha_0} \, \cos{\varphi_k}, \;
\sin{\alpha_0} \, \sin{\varphi_k}, \; -\cos{\alpha_0}) \,.
\end{equation}
Here, $\theta=\pi-\alpha_0$ and $\varphi_k$ are the polar and
azimuthal angles of the initial photon. For a downward pointing
photon (head-on collision), we have $\alpha_0 = 0$. After the
scattering, the four-momentum of the electron is $p'$, and the
scattered photon has energy $\omega'$ and three-momentum
\begin{equation}
{\bm k}'=\omega'(\sin{\theta'} \cos{\varphi'_k}, \; \sin{\theta'}
\sin{\varphi'_k}, \; \cos{\theta'})\,,
\end{equation}
where $\theta'$ and $\varphi'_k$ are the polar and azimuthal
angles of the final photon. Let $\beta$ be the angle between the
vectors ${\bm k}'$ and $(-{\bm k})$. For head-on backscattering,
we have $\beta = 0$. In general,
\begin{equation}
{\bm k}  {\bm k}'= -\omega \; \omega' \; \cos{\beta}\,,
\end{equation}
and
\begin{equation}
\cos{\beta}= \cos{\alpha_0} \, \cos{\theta'} -\sin{\alpha_0} \,
\sin{\theta'} \, \cos{(\varphi_k-\varphi'_k})\,.
\end{equation}
From the on-mass-shell condition of the
scattered electron, we have
$p'^2 = (p+k-k')^2=m_e^2$, and therefore
$k' \cdot (p+k) =k \cdot p$ or
\begin{equation}
\label{omf}
\omega' =
\frac{m_e^2 \, x}{2E(1-v\cos{\theta'}) +
2\omega(1+\cos{\beta})}\,,
\end{equation}
where
\begin{equation}
\label{defx} x = \frac{2 \, k\cdot p}{m_e^2}=\frac{2\omega
E(1+v\cos{\alpha_0})}{m_e^2}\,.
\end{equation}

The $S$ matrix element for plane waves (either plane direct
incoming and outgoing plane waves or plane-wave components of a
twisted photon) is
\begin{align}
\label{Sfidef}
& \langle k',\Lambda', p',\lambda'| S |k, \Lambda, p,\lambda\rangle
\nonumber\\[2ex]
& \qquad =\ii\,(2\pi)^4 \,\delta(p+k-p'-k')\;
\frac{M_{fi}}{4\sqrt{E\,E'\,\omega\,\omega'}}\,,
\end{align}
where the scattering amplitude $M_{fi}$ in the Feynman gauge is
equal to
\begin{subequations}
\begin{eqnarray}
\label{defM}
M_{fi}&=&4\pi\alpha\left(\frac{A}{s-m_e^2}+\frac{B}{u-m_e^2}\right)\,,\\[2ex]
\label{defA} A &=& \bar{u}_{p'\lambda'} \, \hat{e}^*_{k' \Lambda'}
\, \left(\hat p +\hat k +m_e\right)
\hat{e}_{k \Lambda}\, u_{p \lambda}\,,
\\[2ex]
\label{defB} B &=& \bar{u}_{p' \lambda'} \, \hat{e}_{k \Lambda}
\left(\hat{p}' -\hat k +m_e\right) \hat{e}^*_{k' \Lambda'} \, u_{p
\lambda}\,,\\[2ex]
\label{defs} s-m_e^2&=& 2 k\cdot p= m_e^2x\,,\;\;u-m_e^2=
-2k'\cdot p\,.
\end{eqnarray}
\end{subequations}
The bispinors $u_{p \lambda}$ and $u_{p' \lambda'}$ describe the
initial and final electrons with helicities $\lambda$ and
$\lambda'$, and $e_{k \Lambda}$ and $e_{k' \Lambda'}$ are the
polarization vectors of the initial and final photons with
helicities $\Lambda$ and $\Lambda'$. We denote the Feynman dagger
as $\hat p = \gamma^\mu p_\mu$.

\begin{figure}
\begin{center}
\includegraphics[width=0.8\linewidth,angle=0]{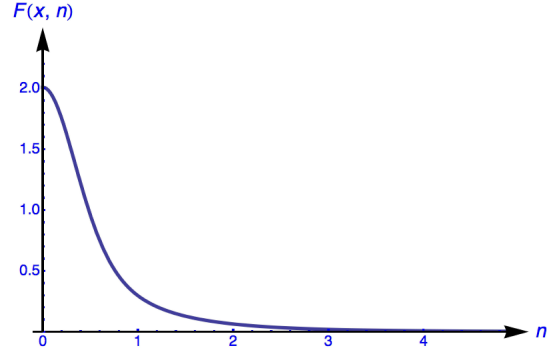}
\caption{\label{fig4} The angular distribution
$\dd\sigma/\dd\Omega'= 2 \alpha^2\, \gamma^2\,m_e^{-2} \, F(x,n)$
of the final photons in the Compton scattering in dependence on $n
= \gamma \, \theta'$. The value $x=0.092$ corresponds to the
VEPP-4M collider.}
\end{center}
\end{figure}

For Compton scattering off incoming ultra-relativistic electrons
($\gamma \gg 1$), the differential cross section has a maximum in
the backscattering region, where the polar angle of the scattered
photon is small, $\theta'\lesssim 1/\gamma$, and the photon
propagates almost along the direction of momentum of the initial
electron. Indeed, for unpolarized electrons, the differential
Compton cross section reads~\cite{BeLiPi1982vol4}
\begin{eqnarray}
\frac{\dd\sigma}{\dd\Omega'}&=&
\frac{2\alpha^2\gamma^2}{m_e^2}\,F(x,n),
\\
F(x,n) &=& \left(\frac{1}{1+x+n^2}\right)^2
\left[\frac{1+x+n^2}{1+n^2} \right.
\nonumber\\[2ex]
& & \left. + \frac{1+n^2}{1+x+n^2}-4 \, \frac{n^2}{(1+n^2)^2}\right]
\,, \nonumber
\end{eqnarray}
where $n=\gamma \, \theta'$, $\dd
\Omega'=\sin{\theta'}\dd{\theta'}\dd\varphi'_{k}$, and
\begin{equation}
\label{defx2} \frac{\omega'}{\omega} = 
\frac{4 \gamma^2}{1+x+n^2} \,.
\end{equation}
In Fig.~\ref{fig4}, we show the angular distribution of the final
photons which is concentrated to the region $n \lesssim 1$. The
value of $x$ used, namely $x=0.092$, corresponds to the collider
parameter $x$ as defined in Eq.~\eqref{defx}, evaluated for the
VEPP-4M collider (Novosibirsk) with $E=5\,{\rm GeV}$ and $\omega
=1.2$~eV. The maximum energy of the final photon is $\omega' =
420\,{\rm MeV}$ for $\theta'=0$. For $n \lesssim 1$, the energy
$\omega'$ of the final photon is independent of the azimuth angle
$\varphi_k$ or $\varphi'_k$,
\begin{equation}
 \label{mainregion}
\omega'=\frac{x}{1+x+(\gamma\theta')^2 }\;E\,.
\end{equation}

To make calculations in the main region more transparent, it is
useful to decompose the scattering amplitude \eqref{defM}
into dominant and negligible items. To do this, in the
$A$ term defined in Eq.~\eqref{defA}, we transpose
$\hat{e}_{k\Lambda}$ and $\hat p +\hat k +m_e$ using the Dirac
equation and obtain
\begin{align}
& \left(\hat p +\hat k +m_e\right)\hat{e}_{k\Lambda}\,
u_{p\lambda} = \; \hat{e}_{k\Lambda} (-\hat p-\hat k +m_e)\,
u_{p\lambda}
\nonumber\\[2ex]
& \quad + 2(e_{k\Lambda} \cdot p)\,u_{p\lambda} = \;
-\hat{e}_{k\Lambda} \hat k\, u_{p\lambda} + 2(e_{k\Lambda} \cdot
p)\,u_{p\lambda} \,.
\end{align}
Thus, $A = A_1 + A_2$ with
\begin{subequations}
\begin{align}
\label{A1} A_1 =& \;  -\bar{u}_{p'\lambda'} \;
\hat{e}_{k'\Lambda'}^* \; \hat{e}_{k\Lambda} \; \hat{k}\,
u_{p\lambda}\,,
\\[2ex]
\label{A2} A_2 =& \; 2 (e_{k\Lambda} \cdot p)\;
\bar{u}_{p'\lambda'} \; \hat{e}_{k'\Lambda'}^*\; u_{p\lambda}\,.
\end{align}
In full analogy, $B=B_1+B_2$ with
\begin{align}
\label{B1} B_1 =& \;  \bar{u}_{p'\lambda'} \; \hat{k} \;
\hat{e}_{k\Lambda} \; \hat{e}_{k' \Lambda'}^* \; u_{p
\lambda}\,,\;\;
\\[2ex]
\label{B2} B_2 =& \; 2 (e_{k\Lambda} \cdot p') \; \bar{u}_{p'
\lambda'} \; \hat{e}_{k'\Lambda'}^*\; u_{p\lambda}\,.
\end{align}
\end{subequations}
The scattering amplitude $M_{fi}$ as defined in Eq.~\eqref{defM}
can thus be written as
\begin{subequations}
\label{Mfi12}
\begin{align}
M_{fi}=& \; M_{1}+M_{2} \,,
\\[2ex]
M_{1} =& \; 4\pi\alpha\left(\frac{A_{1}}{s-m_e^2}+
\frac{B_{1}}{u-m_e^2}\right)\,,
\\[2ex]
M_{2} =& \; 4\pi\alpha\left(\frac{A_{2}}{s-m_e^2}+
\frac{B_{2}}{u-m_e^2}\right)\,.
\end{align}
\end{subequations}
The term $M_{1}$ will be shown to play the dominant role in our
calculation. For further analysis we also introduce three
4-vectors,
\begin{equation}
\label{vec}
\eta^{(\pm)}=\frac{\mp 1}{\sqrt{2}}\,(0,\,1,\pm \ii,\,0)\,,\;\;
\eta^{(z)}=(0,\,0,\,0,\,1) \,.
\end{equation}

%
%
\subsection{Strict backward Compton scattering}

For a head-on collision of a plane-wave photon and a
counter-propagating electron, the electron after scattering moves
in the same direction as before the collision, but with a smaller
energy $E'$,
\begin{equation}
p'=(E',0,0,v'E'), \;\; v'=\frac{|{\bm p}'|} {E'}\,.
\end{equation}
For plane waves, strict backward scattering has the largest
probability (see Fig.~\ref{fig4}), and  this case allows for a
simplified treatment. Indeed, for strict backward geometry, we
have $k'_\perp=k_\perp=0$, and the photon polarization vectors can
be chosen in the form
\begin{equation}
\label{odvec}
e_{k',\pm 1}=e_{k,\mp 1}=\eta^{(\pm)}\,.
\end{equation}
For the considered head-on collision, we have
\begin{subequations}
\begin{align}
e_{k\Lambda}\cdot p=& \; 0\,, \qquad e_{k\Lambda}\cdot p'=0 \,,
\\[2ex]
A_2= & \; 0\,, \qquad B_2=0\,,
\end{align}
\end{subequations}
i.e. the term $M_{2}^{\rm (PW)}$ vanishes for plane-wave strict
backward scattering.

In order to calculate $A_1$ as given by Eq.~\eqref{A1}, it is
useful to represent the expression $\hat{e}_{k'\Lambda'}^* \,
\hat{e}_{k\Lambda}$ as
\begin{align}
2\,\hat{e}_{k'\Lambda'}^* \, \hat{e}_{k\Lambda} =& \; \Lambda
\Lambda'-1 +\left(\Lambda'-\Lambda \right)\Sigma_z
\nonumber\\[2ex]
=& -2\,\left(1+ \Lambda \Sigma_z\right)\,\delta_{\Lambda,
-\Lambda'}\,,
 \label{eepw}
\end{align}
where $\frac 12 {\bm \Sigma}$ is a $4 \times 4$ 
matrix vector measuring
the electron spin. Substituting this expression into $A_1$, we
find
\begin{subequations}
\begin{equation}
\label{A1PW} A^{\rm (PW)}_1 = 8\omega\sqrt{EE'} \,\delta_{\lambda
\lambda'}\,\delta_{\Lambda, -\Lambda'}\,\delta_{2\lambda,\Lambda}
\end{equation}
and, analogously,
\begin{equation}
\label{B1PW} B^{\rm (PW)}_1 =-8\omega\sqrt{EE'} \, \delta_{\lambda
\lambda'}\,\delta_{\Lambda,
-\Lambda'}\,\delta_{2\lambda,-\Lambda}\,.
\end{equation}
\end{subequations}
Further important kinematic relations are
\begin{equation}
s-m_e^2 = m_e^2 \; x \,,\qquad 
u-m_e^2=- \frac{m_e^2 \; x}{x+1} \,.
\end{equation}

As a result, the scattering amplitude for plane-wave strict
backward scattering is
\begin{align}
\label{M1PW} M^{\rm (PW)}_{fi}=& \; M_{1}^{\rm (PW)} = \frac{8\pi
\alpha}{\sqrt{x+1}} \left[\delta_{2\lambda,\Lambda} \right.
\nonumber\\[2ex]
& \left. +(x+1)\delta_{2\lambda,-\Lambda}\right] \,\delta_{\lambda
\lambda'}\,\delta_{\Lambda, -\Lambda'} \,,
\end{align}
with $x=4\omega E/m_e^2$. We emphasize that in head-on
backscattering, the electron does not change its helicity during
scattering ($\lambda'=\lambda$) while the photon does change its
helicity, $\Lambda'=-\Lambda$. All these results are in full
agreement with known properties of ordinary
Compton scattering~\cite{BeLiPi1982vol4,GiKoPoSeTe1984,KoPoSe1998}.

%
%
\section{Compton backscattering of twisted photons}
\label{comptontwisted}

%
%
\subsection{Kinematics}

In the case of a twisted photon, the final $m'$ photon state
$|\varkappa', m', k_z',\Lambda' \rangle$ is a superposition of
plane waves with high energy, consistent with the
general principle of Compton backscattering.
The transverse momentum is conserved,
\begin{equation}
\label{phi} {\bm k}'_\perp={\bm k}_\perp \,,
\end{equation}
as can be seen from the conservation law $k+p=k'+p'$, which
follows from the fact that the transverse components of the vectors
$p$ and $p'$ are equal to zero. The scattering angle is very small,
\begin{equation}
\theta'=\frac{k'_\perp}{\omega'}\lesssim \frac{\omega}{\omega'} =
\frac{1+x +n^2}{4\gamma^2} \approx
\frac{1+x}{4\gamma^2} \,.
\end{equation}
From Eq.~\eqref{phi}, we have $\varphi_k'=\varphi_k$,
and the energy of the scattered photon is
\begin{equation}
\label{energy}
\omega'=\frac{x}{1+x}\, E\,,
\end{equation}
as evident from Eq.~\eqref{mainregion} in the limit $\theta' \to 0$.

In view of the structure of the plane-wave scattering element
recorded in Eq.~\eqref{Sfidef}, the convoluted matrix element for
backward scattering of twisted photons given in
Eq.~(\ref{convol}) takes the form
\begin{align}
\label{tildeSfi} & S^{\rm (TW)}_{fi} =  \ii\,2\pi\, \ii^{m'-m}
\delta(\varkappa-\varkappa') \frac{1}{4\sqrt{EE'\omega\omega'}}
\\[2ex]
& \times
\delta(E+\omega-E'-\omega') \,
\delta(p_z+k_z-p_z'-k_z')
\nonumber\\[2ex]
& \times \int\limits_0^{2\pi}{\ee^{\ii(m-m')\varphi_k}}
\left( M_{1}^{\rm (PWC)} + M_{2}^{\rm (PWC)} \right)%
\dd\varphi_k \,, \nonumber
\end{align}
where we have used the decomposition~\eqref{Mfi12}, and
$M_{1}^{\rm (PWC)}$ and $M_{2}^{\rm (PWC)}$ are the plane-wave
components of the twisted photon scattering matrix element. Note
that $M_{1}^{\rm (PWC)}$ and $M_{2}^{\rm (PWC)}$ are not equal to
their plane-wave counterparts $M_{1}^{\rm (PW)}$ and $M_{2}^{\rm
(PW)}$ because of the nonvanishing conical momentum spread of the
twisted photon.

The admissible values of $m'$ are determined by the
dependence of $A$ and $B$ on $\varphi_k$. In order to carry out
the integration over $\varphi_k$, we have to analyze the
dependence of the polarization vectors $e_{k \Lambda}$ and $e_{k'
\Lambda'}$ on the azimuth angle. To this end, we choose the
polarization vector of the final photon in the scattering
amplitude $M_{fi}$ in the form
\begin{equation}
e_{k'\Lambda'} = -\frac{\Lambda'}{\sqrt{2}}\,
\left( e^{(x')} + \ii \Lambda' \, e^{(y')} \right)\,,
\end{equation}
where the unit vector $e^{(x')}=(0,\,{\bm e}^{(x')})$ is in the
scattering plane, defined by the vectors $\bm p \parallel \bm p'$ and
$\bm k'$, while the unit vector $e^{(y')}$ is orthogonal to it,
\begin{equation}
{\bm e}^{(x')}\, \parallel \; ({\bm p}\times {\bm k}')\times {\bm k}'\,,
\qquad
{\bm e}^{(y')}\, \parallel \; ({\bm p}\times {\bm k}')\,.
\end{equation}
As a result, we have in four-vector component notation
\begin{equation}
\label{e2}
e_{k' \Lambda'} = -\frac{\Lambda'}{\sqrt{2}}\,
\left( \begin{array}{c} 0 \\
\cos{\theta'} \cos{\varphi_k}-\ii\Lambda'\sin{\varphi_k}\\
\cos{\theta'} \sin{\varphi_k}+\ii\Lambda'\cos{\varphi_k}\\
-\sin{\theta'}
\end{array} \right) \,.
\end{equation}
Omitting small terms of the order of $\theta'$, this vector
becomes
\begin{equation}
\label{e2app} e_{k'  \Lambda'} = -
\frac{\Lambda'}{\sqrt{2}}\,(0,\,1,\,\ii\Lambda',\,0)\,
\ee^{-\ii\Lambda'\varphi_k} = \eta^{(\Lambda')}\,
\ee^{-\ii\Lambda'\varphi_k}\,.
\end{equation}
The polarization vector $e_{k \Lambda}$ of a conical
component of the initial twisted photon (as a function of
$\varphi_k$) is obtained by setting $\theta'=\pi-\alpha_0$ in
$e_{k'  \Lambda'}$ and reads
\begin{equation}
\label{ephi} e_{k  \Lambda}=\frac{\Lambda}{\sqrt{2}}\,
\left( \begin{array}{c} 0\\
\cos{\alpha_0} \cos{\varphi_k}+\ii\Lambda\sin{\varphi_k}\\
\cos{\alpha_0} \sin{\varphi_k}-\ii\Lambda\cos{\varphi_k}\\
\sin{\alpha_0}
\end{array} \right) \,.
\end{equation}
Using the 4-vectors defined in Eq.~\eqref{vec}, we may write it in
the form
\begin{align}
\label{e1decomp} e_{k \Lambda}=& \; \eta^{(-\Lambda)} \, \ee^{\ii
\Lambda \varphi_k}\, \cos^2\left(\frac{\alpha_0}{2}\right) +
\eta^{(\Lambda)} \, \ee^{-\ii\Lambda \varphi_k}\,
\sin^2\left(\frac{\alpha_0}{2}\right) \nonumber
\\
& \; + \frac{\Lambda}{\sqrt{2}}\,\eta^{(z)}\, \sin{\alpha_0}\,.
\end{align}
With the help of Eq.~\eqref{int}, we can also write the Fourier
transform of the product
$e_{k  \Lambda}\, a_{\varkappa m}({\bm k}_\perp)$,
which is still a 4-vector, as
\begin{align}
\label{psi} I_{\varkappa m \Lambda} =& \; \int e_{k \Lambda} \,
a_{\varkappa m}({\bm k}_\perp)\, \ee^{\ii \, {\bm k}_\perp  {\bm
r}} \frac{\dd^2k_\perp}{(2\pi)^2}
\\
=& \; \ii^{\Lambda} \; \eta^{(-\Lambda)} \;
\psi_{\varkappa, m+\Lambda}(r,\varphi_r)\,
\cos^2\left( \frac{\alpha_0}{2}\right)
\nonumber \\
-& \;\ii^{\Lambda} \; \eta^{(\Lambda)} \; \psi_{\varkappa,
m-\Lambda}(r,\varphi_r) \, \sin^2 \left( \frac{\alpha_0}{2}
\right)
\nonumber \\
+& \;\frac{\Lambda}{\sqrt{2}}\, \eta^{(z)}\, \psi_{\varkappa
m}(r,\varphi_r)\, \sin{\alpha_0} \,, \nonumber
\end{align}
recalling that the function $\psi_{\varkappa m}(r,\varphi_r)$ is given in
Eq.~\eqref{defPsi}. With this formula for $I_{\varkappa m \Lambda}$,
we can write the twisted photon
vector potential~(\ref{twistedwave}) as
\begin{equation}
\label{Amu} 
{\cal A}^\mu_{\varkappa m k_z\Lambda}(r,\varphi_r,z,t) =\;
\sqrt{4\pi} \;
\frac{\ee^{-\ii (\omega t - k_z z)}}{\sqrt{2\omega}}\; 
I^\mu_{\varkappa m \Lambda}\,.
\end{equation}
This 4-vector potential corresponds to the initial twisted photon
state $|\varkappa, m, k_z,\Lambda \rangle$, which describes a
superposition of states with projections of the orbital angular
momentum onto the $z$ axis equal to $m$ and $m\pm \Lambda$. If the
angle $\alpha_0$ becomes small, we have
\begin{equation}
|\varkappa, m, k_z,\Lambda \rangle \propto
\eta^{(-\Lambda)} \; \psi_{\varkappa, m+\Lambda}(r,\varphi_r) \,,
\end{equation}
and the projection $m+\Lambda$ becomes dominant. For the final twisted
photon with its small angle $\theta'$, we have
\begin{equation}
|\varkappa', m', k_z',\Lambda' \rangle \propto
\eta^{(\Lambda')} \; \psi_{\varkappa', m'-\Lambda'}(r,\varphi_r),
\end{equation}
and, therefore, the projection $m'-\Lambda'$ is dominant.

%
%
\subsection{Main $\maybebm{S}$ matrix contribution for twisted photons}

For backward scattering of twisted photons, the incoming and
outgoing polarization vectors are given by Eqs.~\eqref{ephi}
and~\eqref{e2app}, respectively. Investigating the contribution
from $A_1$ given by \eqref{A1}, we write $\hat{e}_{k'\Lambda'}^*
\, \hat{e}_{k\Lambda}$ in the form
\begin{align}
2\hat{e}_{k'\Lambda'}^* \; \hat{e}_{k\Lambda} &= -(1-\Lambda
\Lambda'\cos{\alpha_0})\,(1-\Lambda' \Sigma_z)
\nonumber\\
&\quad +\Lambda \, \left(\Sigma_x-\ii \, \Lambda' \, \Sigma_y
\right)\, \ee^{\ii \Lambda' \varphi_k}\,.
\label{eetw}
\end{align}
Substituting this expression in $A_1$, we find
\begin{subequations}
\begin{align}
\label{A1PWC} A_1^{\rm (PWC)} =
 & \; 2 \omega \, \sqrt{EE'}\, \left[(1-\Lambda \,
\Lambda'\cos{\alpha_0}) \, (1+\cos{\alpha_0}) \right.
\nonumber\\
& \; \left. +2\lambda \,\Lambda \, \sin^2{\alpha_0}
\right]\,\delta_{\lambda\lambda'}\,\delta_{2\lambda, -\Lambda'} \,.
\end{align}
Analogously,
\begin{align}
\label{B1PWC} B_1^{\rm (PWC)} = &-2 \omega \, \sqrt{EE'}\,
\left[(1-\Lambda \, \Lambda'\cos{\alpha_0}) \, (1+\cos{\alpha_0})
\right.
\nonumber\\
& \; \left. -2\lambda \,\Lambda \, \sin^2{\alpha_0}
\right]\,\delta_{\lambda\lambda'}\,\delta_{2\lambda,\Lambda'}\,.
\end{align}
\end{subequations}
For $\alpha_0 = 0$, the expressions $A_1^{\rm (PWC)}$ and
$B_1^{\rm (PWC)}$ for twisted backscattering become proportional
to the Kronecker delta of the helicities $\delta_{\Lambda, -
\Lambda'}$, and Eqs.~\eqref{A1PWC} and \eqref{B1PWC} coincide with
Eqs.~\eqref{A1PW} and \eqref{B1PW}, respectively. However, for a
twisted photon, $\alpha_0 = \arctan[ \varkappa/(-k_z) ] \neq 0$
and the value $\Lambda'=\Lambda$ is possible. Nevertheless, the
corresponding probability for $\Lambda'=\Lambda$ is small for
small values of the colliding angle $\alpha_0$ because in this
case, $A_1^{\rm (PWC)}\propto \sin^2(\alpha_0/2)$ and $B_1^{\rm
(PWC)}\propto \sin^2(\alpha_0/2)$.

None of the quantities $A_1^{\rm (PWC)}$,
$B_1^{\rm (PWC)}$ nor $M_1^{\rm (PWC)}$ depend on $\varphi_k$. This could
be expected because in the considered case, the colliding plane
(determined by the vectors ${\bm p}$ and ${\bm k}_\perp$)
coincides with the scattering plane (determined by the vectors
${\bm p}$ and ${\bm k}'_\perp={\bm k}_\perp$). The integral over $
\varphi_k$ in Eq.~\eqref{tildeSfi} is trivial, and we thus have
$m'=m$ in the term $M_1^{\rm (PWC)}$. Finally,
\begin{equation}
\int\limits_0^{2\pi}{\ee^{\ii(m-m')\varphi_k}} M_1^{\rm (PWC)}
\dd\varphi_k = 2 \pi \, \delta_{m m'} \, M_1^{\rm (PWC)} \,,
\end{equation}
with
\begin{equation}
\label{M1} M_1^{\rm (PWC)} = 4\pi\alpha \; \left(\frac{A_1^{\rm
(PWC)}}{s-m_e^2} + \frac{B_1^{\rm (PWC)}}{u-m_e^2}\right) \,.
\end{equation}
The quantities $A_1^{\rm (PWC)}$ and $B_1^{\rm (PWC)}$ are given
in Eqs.~\eqref{A1PWC} and \eqref{B1PWC}.

Collecting all prefactors, we can establish the following
structure for the $S$ matrix element~\eqref{convol} for strict
backward Compton scattering of twisted photons,
\begin{align}
\label{mainres} S^{\rm (TW)}_{fi} = & \; 
\langle \varkappa', m', k'_z, \Lambda'; p', \lambda' | S | 
\varkappa, m, k_z,\Lambda; p, \lambda \rangle
\nonumber\\[2ex]
= & \;  \ii \, (2 \pi)^2\,\delta_{m m'} \,
\delta(\varkappa-\varkappa') \,\delta(E+\omega-E'-\omega')
\nonumber\\
& \; \times \; \delta(p_z+k_z-p_z'-k_z') \frac{M_{1}^{\rm
(PWC)}}{4\sqrt{EE'\omega\omega'}} \,.
\end{align}
%

%
%
\subsection{Neglected contribution $\maybebm{M_{2}^{\rm (PWC)}}$ }

Let us consider the terms $A_{2}^{\rm (PWC)}$ and $B_{2}^{\rm
(PWC)}$ which enter the expression \eqref{Mfi12} for $M_{2}^{\rm
(PWC)}$. These items do not vanish, as for plane-wave strict
backscattering, but there are large cancellations among them. We
write  $M_{2}^{\rm (PWC)}$ as follows,
\begin{align}
M_{2}^{\rm (PWC)} =& \; 4\pi\alpha\; \bar{u}_{p'\lambda'} \,
\hat{e}_{k'\Lambda'}^*\,u_{p\lambda}\,\left(\frac{e_{k\Lambda}\cdot
p}{k\cdot p}-\frac{e_{k\Lambda}\cdot p'}{k\cdot  p'}\right)
\nonumber\\[2ex]
=& \; -4\pi\alpha\;\bar{u}_{p'\lambda'} \,
\hat{e}_{k'\Lambda'}^*\,u_{p\lambda} \,
\frac{(e_{k\Lambda})_z}{\omega}\,\epsilon\,,
\end{align}
where the quantity
\begin{equation}
\epsilon=\omega\,\left(\frac{p_z}{k\cdot p} - \frac{p'_z}{k\cdot
p'}\right)
\end{equation}
is very small due to mutual cancellations,
\begin{align}
 \label{cancel}
\epsilon=& \;
\frac{v}{1+v\cos{\alpha_0}}-\frac{v'}{1+v'\cos{\alpha_0}}
\nonumber\\[2ex]
=& \; \frac{m_e^2}{2E^2}\,\frac{x(x+2)}{(1+\cos{\alpha_0})^2}\ll
1\,.
\end{align}
Finally,
\begin{equation}
\label{As}
\bar{u}_{p'\lambda'} \; e_{k'\Lambda'}^* \; u_{p\lambda} =
\sqrt{2} \, \ee^{\ii \, \Lambda' \varphi_k}\;
\frac{m_e \, \omega'}{\sqrt{EE'}}\,
\delta_{\lambda, -\lambda'}\,
\delta_{2\lambda, -\Lambda'}\,.
\end{equation}
The main contribution in this result is given by the
transverse component of the vector ${e}_{k'\Lambda'}$,
while the longitudinal component $\left(e_{k'\Lambda'}\right)_{z}$ gives
only a small contribution,
\begin{equation}
|\bar{u}_{p'\lambda'} \; \left(\hat{e}_{k'\Lambda'}\right)_{z}^*
\; u_{p\lambda}|\lesssim \frac{\omega}{\omega'} \, \sqrt{EE'} \,,
\end{equation}
since $|\left(e_{k'\Lambda'}\right)_{z}|\lesssim \theta'$. In view
of Eq.~\eqref{cancel}, we have  $|M_{2}^{\rm (PWC)}| \ll
|M_{1}^{\rm (PWC)}|$ due to the mutual cancellations of large
contributions from $s$ and $u$ channels. We can thus neglect
$M_{2}^{\rm (PWC)}$ in strict backward scattering.

%
%
\subsection{Averaged cross section for twisted photons}
\label{avcrossX}

We base our considerations in this section on the approach
described in Ref.~\cite{BeLiPi1982vol4}, taking into account the
necessary modifications for twisted photons. Compton scattering
needs to be considered in a large but finite space-time volume
$T\,V$, with time duration $T$ and spatial volume $V$. The
normalization of plane-wave particles with $V=L_xL_yL_z$ as well
as twisted photons with $V=\pi R^2L_z$ is discussed in detail in
the Appendices~\ref{appa} and~\ref{appb}.

The $S$ matrix element for the Compton scattering of plane-wave
particles has to be written as [see Eq.~\eqref{Sfidef}]
\begin{equation}
S^{\rm (PW)}_{fi}=\ii\,(2\pi)^4 \,\delta(p+k-p'-k')\; \frac{M^{\rm
(PW)}_{fi}}{4\sqrt{E\,E'\,\omega\,\omega'}}\,\frac{1}{V^2}\,,
\end{equation}
where the last factor $1/V^2$ takes into account the normalization
for plane-wave electrons and photons. We consider the head-on
collisions of the initial particles when current densities are
$j_z^{(e)}=v/V$ and $j_z^{(\gamma)}=-1/V$
(see Appendix~\ref{appa}). The corresponding cross section 
for Compton scattering is
equal to the probability of the process
\begin{equation}
\dd W_{fi}=\left|S^{\rm (PW)}_{fi}\right|^2\dd n_e \, \dd n_\gamma \,,
\end{equation}
divided over the time $T$ and the
current density $j_z$ of the colliding particles,
\begin{equation}
\dd \sigma=\frac{\dd W_{fi}}{T
j_z}\,,\;\;j_z=j_z^{(e)}-j_z^{(\gamma)}=\frac{v+1}{V}\,.
\end{equation}
Here $\dd n_e=V\dd^3 p'/(2\pi)^3$ and $\dd n_\gamma=V\dd^3
k'/(2\pi)^3$ are the number of states for the final electron and
final photon in the given phase-space volumes. The obtained quantity
\begin{equation}
\label{crpw} \dd \sigma^{\rm (PW)} =
\frac{\delta(E+\omega-E'-\omega')}{16(2\pi)^2(v+1)} \,
\frac{\left|M_{fi}^{\rm (PW)}\right|^2}{EE'\omega\omega'}\, \dd^3
p'\,,
\end{equation}
does not depend on $T$, $V$ and neither on the 
coordinates in the transverse plane.

Let us now consider the Compton scattering with twisted
photons assuming that the propagation axis of the initial photons
is antiparallel to the momentum of the initial electron. In such a
case, the current density of initial photon, given by
Eq.~\eqref{dentjzr}, depends on the radial variable~$r$, i.e., on
the distance from the central symmetry axis of the twisted photon.
This implies that the usual notion of a cross section, which is
normalized to an incoming plane-wave flux of particles, uniform in
the plane normal to the propagation vector of the incoming
particles, actually cannot be applied in this case. Therefore, we
need some kind of generalization of the usual notion of a cross
section. In some aspect, this situation resembles the case of
processes with large impact parameters mentioned in
Sec.~\ref{intro}.

In order to characterize the discussed process quantitatively, we suggest
to use the averaged current density
derived in Sec.~\ref{appb} below [see Eq.~\eqref{curr}] and define an {\it
averaged cross section} as
\begin{equation}
 \label{cr}
\dd \sigma_{\rm av} =\frac{\left|S^{\rm (TW)}_{fi}\right|^2}{T \,
\langle j_z\rangle} \dd n_e \; \dd n_\gamma,\,
\end{equation}
where
\begin{align}
\label{crdop} \dd n_e =& \; \frac{V\dd^3 p'}{(2\pi)^3}, \;\; \dd
n_\gamma= \frac{R \, \dd \varkappa'}{\pi} \, \frac{L_z \, \dd
k_z'}{2\pi} \,,
\nonumber\\[2ex]
\langle j_z\rangle=& \;j_z^{(e)}-\langle j^{(\gamma)}_z\rangle=
\frac{v+\cos{\alpha_0}}{V}\,,
\end{align}
and the $S$ matrix element for strict backward Compton scattering
of twisted photons has to be written as [compare
with Eq.~\eqref{mainres}]
\begin{align}
& \label{Stw} S^{\rm (TW)}_{fi} = \ii \, (2 \pi)^2\,
\delta_{m m'} \,
\delta(\varkappa-\varkappa') \, \delta(E+\omega-E'-\omega')
\nonumber\\
& \; \times \; \delta(p_z+k_z-p_z'-k_z') \frac{M_{1}^{\rm
(PWC)}}{4\sqrt{EE'\omega\omega'}} \frac{\pi}{VRL_z}\,.
\end{align}
Here, the last factor $\pi/(VRL_z)$ takes into account the
normalization for the plane-wave electrons and twisted photons.

In the standard approach~\cite{BeLiPi1982vol4}, the squared delta function
$\delta(E+\omega-E'-\omega')\equiv \delta(\epsilon)$ is understood
as
\begin{equation}
\left[ \delta(\epsilon) \right]^2 = \delta(\epsilon)\,\frac{1}{2
\pi} \int_{-T/2}^{T/2} \ee^{\ii \epsilon t}\dd t= \frac{T}{2 \pi}
\delta(\epsilon)\,.
\end{equation}
Analogously,
\begin{equation}
\left[\delta(p_z+k_z-p_z'-k_z')\right]^2 = \frac{L_z}{2
\pi}\delta(p_z+k_z-p_z'-k_z')\,.
\end{equation}
For the last delta function we obtain, in the limit of a large
radial dimension $R$ of the normalization volume for the twisted state,
\begin{subequations}
\begin{equation}
|\delta(\varkappa-\varkappa')|^2= \frac{R}{\pi}
\delta(\varkappa-\varkappa') \,.
\end{equation}
This result can be derived
with the help of the identities~\eqref{identJJ} and \eqref{Jmsqrt},
in the following form
\begin{equation}
\delta(\varkappa-\varkappa')\big|_{\varkappa\to\varkappa'}
= \int_0^{R} \dd r \,\varkappa \, r \, J^2_m(\varkappa r)
= \frac{R}{\pi}\,.
\end{equation}
\end{subequations}
As a result, all factors $T$, $V$, $R$, and $L_z$ disappear in the averaged
cross section. After integration over $k_z'$ and $\varkappa'$
we obtain, for strict backward scattering,
\begin{equation}
\label{avcr} \dd \sigma_{\rm av}^{\rm (TW)}
=\delta_{mm'}\,\frac{\delta(E+\omega-E'-\omega')} {16 (2 \pi)^2
(v+\cos{\alpha_0})} \, \frac{\left|M_{1}^{\rm
(PWC)}\right|^2}{EE'\omega\omega'}\,\dd^3 p' \,.
\end{equation}
Here, $M_{1}^{\rm (PWC)}$ is given in Eq.~\eqref{M1} and
corresponds to the amplitude for the case where an initial
plane-wave photon collides with the initial electron at a given
collision angle $\alpha_0\neq 0$ (not a head-on collision!). The
result~\eqref{avcr} follows naturally
because the initial photon state for a
defined quantum number $m$ is nothing else but a superposition of
plane waves with the same absolute value of their transverse
momenta.

%
%
\section{Conclusions}
\label{conclu}

In this paper, we have investigated the scattering of a twisted
laser photon by an ultra-relativistic incoming electron, in the
Compton backscattering geometry.
The electron is described as an incoming plane wave
(in contrast to Compton scattering from bound 
electrons~\cite{SuBePiPr1991,BeSuPiPr1993}),
but the wave function of the incoming photon is nontrivial.
A twisted photon is a state with
a definite orbital angular momentum projection $m$ on the
propagation axis. We put special emphasis on the particular but
important case of backward Compton scattering and perform a
detailed calculation for this case. As a result, we prove the
principal possibility to create high-energy photons with high
energy and large orbital angular momenta projection.

From the experimental point of view, strict backscattering can be
realized by detecting the electrons scattered at zero angle.  A
technique for the registration of electrons scattered at small
(even zero) angles after the loss of energy in the Compton process
is implemented, for example, in the device for backscattered
Compton photons installed on the VEPP-4M collider
(Novosibirsk)~\cite{NeTuSh2004}.

The main result of the paper is contained in Eqs.~\eqref{mainres}
and~\eqref{avcr}, which give the amplitude and  the averaged cross
section for the transition in which an incoming twisted photon
with quantum numbers $|\varkappa, m, k_z, \Lambda \rangle$ is
scattered into a twisted photon state $|\varkappa', m', k'_z,
\Lambda' \rangle$. According to Eq.~\eqref{mainres}, the magnetic
quantum number $m'=m$ and  the conical momentum spread $\varkappa'
= \varkappa$ are preserved, but the energy of the final twisted
photon is increased dramatically:
$\omega'/\omega \sim \gamma^2 \gg 1$.
This implies that the conical angle $\theta'$ of the scattered twisted
photon is very small, with
$\theta' \approx \varkappa'/\omega' \sim 1/\gamma^2$.

One of the most interesting applications of high-energy twisted
photons would concern the irradiation of heavy nuclei. Indeed,
there are plans at GSI Darmstadt~\cite{HITRAP} to slow down and
investigate heavy ions in Penning traps. Giant dipole resonances
and following fission of nuclei have been the subject of
the investigations~\cite{CaEtAl1980,BuEtAl1991,HaWo2001}. Typically,
giant dipole resonances are in the range of $\sim$~10--30~MeV.
Irradiation of nuclei by twisted photons with frequencies below
the first resonance might reveal new and fundamental insight into
the dynamics of a fast rotating quantum many-body system.

Another interesting application would be concerned
with the direct excitation of atomic or ionic ground states
into high-lying, circular Rydberg states.
The orbital angular momentum would in this case act
as a ``quantum kicker'' with excitation energies
(for heavy ions) in the range of several hundred~eV.

\section*{Acknowledgments}

We are grateful to I.~Ginzburg, D.~Ivanov, I.~Ivanov, G.~Kotkin,
A.~Milstein, N.~Muchnoi, O.~Nachtmann, V.~Telnov, A.~Voitkiv,
V.~Zelevinsky and V.~Zhilich for useful discussions. U.D.J.~and
V.G.S.~acknowledge support from the Missouri Research Board. In
addition, this research has been supported by the National Science
Foundation (U.D.J.) and by the Russian Foundation for Basic
Research (grants 09-02-00263 and NSh-3810.2010.2, V.G.S.).

\appendix

%
%

%
%
\section{Normalization for plane-wave particles}
\label{appa}

We first discuss a scalar particle. For the calculation of the cross
section, it will be convenient to consider a field in the large
but finite volume $V=L_x\,L_y\,L_z$. It is well
known~\cite{BeLiPi1982vol4} that for scalar particles with zero mass,
the appropriate plane-wave solution is of the form
\begin{equation}
\label{pl-wave} \psi_{\bm k}(x) = \frac{\ee^{-ik \cdot
x}}{\sqrt{2\omega V}}\,,\;\; k \cdot x =\omega \, t-{\bm k}{\bm
r},\;\; \omega=|{\bm k}| \,.
\end{equation}
This corresponds to the constant density
\begin{equation}
\rho =\ii\, \psi^*_{\bm k}(x) \, \partial_t\psi_{\bm k}(x) +
\mbox{c.c.} = \frac{1}{V} \,,
 \label{den}
\end{equation}
i.e., to one particle in the volume $V$. The current density for
the initial particle $j_z$ then is also constant and reads
\begin{equation}
 \label{cd}
j_z=\frac{k_z/\omega}{V}\,.
\end{equation}

We can calculate the number of admissible states by integrating
the infinitesimal phase-space volume $k_z \dd z$ from $z=-L_z/2$
to $z = L_z/2$ and postulating that there is one available quantum
state per phase-space volume $2\pi$,
\begin{equation}
 \label{adiabz}
n_z = \int \frac{k_z\dd z}{2 \pi} = L_z \, \frac{k_z}{2\pi}\,.
\end{equation}
The number of available states for the final particle in the
interval $\dd k'_z$ is, therefore,
\begin{equation}
\dd n_z = L_z \, \frac{\dd k_z'}{2 \pi}\,.
\end{equation}
In three dimensions, the number of states for the final particle
thus is found to be
\begin{equation}
\label{number}
\dd n = \dd n_x \, \dd n_y \, \dd n_z =
V \, \frac{\dd^3 k'}{(2\pi)^3}\,.
\end{equation}

A plane-wave vector photon is normalized as follows. 
Let us now consider the vector plane-wave photon. It is convenient
to use the Coulomb gauge in which the photon field is:
\begin{equation}
A^\mu(x)=(0,\,{\bm A}(x)),\;\;\; {\bm \nabla}  {\bm A}(x)=0\,.
\end{equation}
Then the electric ${\bm E}$ and magnetic ${\bm B}$ fields
are determined by the vector potential only,
\begin{equation}
{\bm E} = -
\frac{\partial}{\partial t} {\bm A}(x),\; \qquad
{\bm B}= \bm\nabla \times {\bm A}(x) \,,
\end{equation}
and the field operator can be presented in the form
\begin{equation}
\check{\bm A}(x) = \sum_{{\bm k}\Lambda} \left[\check{a}_{{\bm
k}\Lambda} \, {\bm A}_{{\bm k}\Lambda}(x) + \check{a}^+_{{\bm
k}\Lambda} \, {\bm A}^+_{{\bm k}\Lambda}(x) \right]\,,
\end{equation}
where $\check{a}^+_{{\bm k}\Lambda}$ ($\check{a}_{{\bm k}\Lambda}
$) is the operator for creation (annihilation) of the photon with
momentum ${\bm k}$ and helicity $\Lambda$. The plane-wave solution
now is
\begin{equation}
{\bm A}_{{\bm k}\Lambda}(x) = \sqrt{4\pi}\, \frac{\ee^{-\ii k\cdot
x}}{\sqrt{2\omega V}}\, {\bm e}_{{\bm k}\Lambda} \,,
\end{equation}
and the polarization vectors satisfies the conditions:
\begin{equation}
{\bm e}_{{\bm k}\Lambda}{\bm k}=0,\quad
{\bm e}^*_{{\bm k}\Lambda'}{\bm e}_{{\bm k}\Lambda} =
\delta_{\Lambda \Lambda'}\,.
\end{equation}
After quantization, the energy of the photon field (in Gaussian
units)
\begin{align}
\label{energy1} E =& \; \frac{1}{8\pi} \int\limits_V \dd^3r \,
\left( {\bm E}^2 + {\bm B}^2 \right)
\nonumber\\[2ex]
=& \; \frac{1}{8\pi} \int\limits_V \dd^3 r \, \left[ \left(
\frac{\partial}{\partial t}  {\bm A} \right)^2 - {\bm A}  \left(
\frac{\partial^2}{\partial t^2} {\bm A} \right) \right]
\end{align}
transforms to the Hamiltonian
\begin{equation}
\check H = \sum_{{\bm k}\Lambda} \omega \, \check{n}_{{\bm
k}\Lambda} \,,
\end{equation}
where $\check{n}_{{\bm k}\Lambda} = \check{a}^+_{\bm k \Lambda} \,
\check{a}_{\bm k \Lambda}$ is the operator for the number of
particles. This form of Hamiltonian corresponds to a normalization
to one photon in the volume $V$ and, therefore, to
the density given in Eq.~\eqref{den}, to the
current density as indicated in Eq.~\eqref{cd},
and to the number of states for the final photon
with helicity $\Lambda'$ given in Eq.~\eqref{number}, respectively.

%
%
\section{Normalization for twisted particles}
\label{appb}

A twisted scalar particle can be discussed as follows.
Let us now consider a twisted
scalar field in a large but finite cylindrical volume $V= (\pi
R^2) \, L_z$ and let the analog of the plane wave be
\begin{equation}
\label{tw-wave} \psi_{\nu}(x) = N\,\frac{\ee^{-\ii\,(\omega t- k_z
z)}}{\sqrt{2\omega}}\, \frac{\ee^{\ii m \varphi_r}}{\sqrt{2\pi}}\,
\sqrt{\varkappa}\, J_m(\varkappa r)\,,
\end{equation}
where the energy $\omega= \sqrt{\varkappa^2+k_z^2}$ and the
quantum numbers for the twisted state are
\begin{equation}
\nu =\{\varkappa,m,k_z\}.
\end{equation}
We can find the factor $N$ from the normalization
of the one-particle state to the volume $V$,
\begin{equation}
\label{normV} \int\limits_V\rho(\bm r) \, \dd^3r= 1\,,
\end{equation}
where the density
\begin{equation}
\rho = \ii\,\psi^*_{\nu}(x)\, \frac{\partial}{\partial
t}\psi_{\nu}(x) + \mbox{c.c.} = N^2
\frac{\varkappa}{2\pi}\,J^2_m(\varkappa r)
 \label{dent}
\end{equation}
as well as the current density
\begin{equation}
j_z = -\ii\psi^*_{\nu}(x)\frac{\partial}{\partial z}\psi_{\nu}(x)
+ \mbox{c.c.} = \frac{k_z}{\omega}N^2
\frac{\varkappa}{2\pi}J^2_m(\varkappa r),
 \label{dentjzr}
\end{equation}
now depend on the radial variable $r$. This yields the condition
\begin{equation}
1=N^2 L_z \int_0^{R} J_m^2(\varkappa r)\, \varkappa r \,
\dd r\,.
\end{equation}
The main contribution to this integral comes from large radial
arguments $r\sim R\gg 1/\varkappa$, where we can use the
asymptotics of the Bessel function and find
\begin{equation}
\label{Jmsqrt}
\int\limits_0^{R} J_m^2(\varkappa r)\,\varkappa \, r \, \dd r 
\approx
\frac{2}{\pi} \int\limits_0^{R} 
\cos^2{\left(\varkappa \, r - \frac{m\pi}{2} - 
\frac{\pi}{4} \right)}\,\dd r \approx \frac{R}{\pi} \,.
\end{equation}
The normalization prefactor $N$ is thus given by
\begin{equation}
\label{Nt}
N=\sqrt{\frac{\pi}{R \, L_z}}\,.
\end{equation}
Below we also will use the density for the initial twisted
particle averaged in the transverse plane
\begin{equation}
\label{avdent}
\langle\rho\rangle \equiv \int\limits_0^R
\rho(r)\,\frac{2\pi\,r \; \dd r}{\pi \, R^2} 
= \frac{1}{\pi \, R^2 \, L_z} 
= \frac{1}{V} \,,
\end{equation}
The ``wave functions'' (vector potentials) of the 
twisted photons are normalized to a Dirac $\delta$ in the 
continuum case. Here, we convert this normalization 
to that for a finite volume $\pi \, R^2 \, L_z$.
The asymptotics of the Bessel function for large 
argument imply that as $R\to\infty$, the oscillations/fluctuations of the 
Bessel function average out, and we can assign an average 
incoming current density 
\begin{equation}
\label{curr}
\langle j_z \rangle = \frac{k_z/\omega}{V} =
-\frac{\cos{\alpha_0}}{V}\,.
\end{equation}
to the twisted photons.
In analogy to Eq.~\eqref{adiabz}, we now count the available
states in the radial variable by evaluating the adiabatic
invariant with respect to the variable~$r$,
\begin{subequations}
\begin{align}
n_r =& \; 2\int\limits_{m/\varkappa}^R \frac{k_r(r) \, \dd r}{2 \pi}
\approx \frac{R \, \varkappa}{\pi}\,,
\\[2ex]
k_r(r) =& \; \sqrt{\varkappa^2- \left( \frac{m}{r} \right)^2}
\approx \varkappa\,.
\end{align}
\end{subequations}
We find the corresponding number of states per interval
$\dd\varkappa$ as
\begin{equation}
\dd n_r = \frac{R \, \dd \varkappa}{\pi}\,.
\end{equation}
As a result, the number of states for the final twisted particle
with a defined quantum number $m'$ in the interval $\dd
\varkappa'\dd k'_z$ is
\begin{equation}
 \label{numvec}
\dd n_t = \dd n_r \, \dd n_z = \frac{R \; \dd \varkappa'}{\pi} \,
\frac{L_z \; \dd k_z'}{2\pi}\,.
\end{equation}
We can thus count all available states for the twisted scalar
particle in a given large cylindrical normalization volume with
radial dimension $R$ and $z$ dimension $L_z$  by
integrating over $\dd n_t$ and summing over all $m'$.

The normalization for the twisted photon is slightly more complicated.
We use the field operator $\check A(x)=(0,\,\check{\bm
A}(x))$ for the 4-vector potential in the form
\begin{equation}
\check{\bm A}(x) = \sum_{\nu} \left[\check{a}_{\nu} \,{\bm
A}_{\nu}(x) + \check{a}^+_{\nu} \, {\bm A}^+_{\nu}(x) \right]\,,
\end{equation}
where the multi-index $\nu$ gives the quantum numbers of the state as
\begin{equation}
 \label{nugamma}
{\nu} =\{\varkappa,m,k_z,\Lambda \}\,.
\end{equation}
The appropriate analog of the plane wave is
\begin{equation}
{\bm A}_{\nu}(x) = \sqrt{4\pi}\,N\, \frac{\ee^{-\ii (\omega t -
k_z\,z)}}{\sqrt{2\omega}}\, {\bm I}_{\varkappa m
\Lambda}(r,\varphi_r) \,,
\end{equation}
where the normalization factor $N$ is given in~\eqref{Nt}, and the
vector ${\bm I}_{\varkappa m \Lambda}(r,\varphi_r)$ is given
in~\eqref{psi}. Using the equalities
\begin{align}
 \label{con}
&\int_V {\bm A}_{\nu}(x){\bm A}^*_{\nu'}(x)\,\dd^3r  =
\frac{2\pi}{\omega} \delta_{\nu\nu'}\,,
\nonumber\\[2ex]
&\int_V {\bm A}_{\nu}(x){\bm A}_{\nu'}(x)\,\dd^3r \propto
\delta_{\varkappa\varkappa'}\delta_{k_z,-k'_z}
\end{align}
we can transform the energy of the photon field \eqref{energy1} to
the following Hamiltonian, written in the basis of twisted photon
wave functions,
\begin{equation}
\check H = \sum_{\nu} \; \omega \; \check n_{\nu} \,, \;\; \check
n_{\nu} = \check{a}^+_{\nu} \check{a}_{\nu}\,,
\end{equation}
This form of the Hamiltonian corresponds to a normalization to one
photon in the volume $V$ and, therefore, to the radially
averaged density given in Eq.~\eqref{avdent},
to the averaged current density given
in~\eqref{curr} and to the number of states for the final photon
with defined quantum numbers $m'$ and $\Lambda'$ given in
Eq.~\eqref{numvec}.

\end{document}